# Fast determination of phases in Li$_x$FePO$_4$ using low losses in electron energy-loss spectroscopy

P. Moreau[1, a)], V. Mauchamp[2], F. Pailloux[2] and F. Boucher[1]

[1]*Institut des Matériaux Jean Rouxel (IMN), Université de Nantes, CNRS, 2, rue de la Houssinière, BP32229, 44322 Nantes Cedex 3, France*

[2]*PhyMAT (ex LMP) SP2MI – Boulevard 3, Téléport 2 – BP 30179 - 86962 Futuroscope, Chasseneuil Cedex, France*

Experimental valence electron energy loss spectra, obtained on different phases of Li$_x$FePO$_4$ are analyzed with first principles calculations based on density functional theory. In the 4-7 eV range, a large peak is identified in the FePO$_4$ spectrum, but is absent in LiFePO$_4$, which allows the easy formation of energy filtered images. The intensity of this peak, non sensitive to the precise orientation of the crystal, is large enough to rapidly determine existing phases in the sample and permit future dynamical studies.

a) Electronic mail: Philippe.Moreau@cnrs-imn.fr

Rechargeable lithium batteries are of considerable technological interest in the field of portable electronic devices and will soon be the standard electrical storage system in hybrid or all-electric vehicles.[1] Among positive electrode materials, lithium iron phosphate $LiFePO_4$ is one of the most promising. It is cheap, non toxic and shows excellent cycling properties when particles are coated with a carbon layer. Even if this coating and the utilization of nanosized particles clearly compensates for its intrinsic poor electronic conductivity, the precise microscopic processes at work inside the crystals are still being debated.[2-5] Given the crucial role of defects, interfaces, surfaces and local amorphous regions, the relevant phenomena are at the nanometer scale. They also involve very short time scales, due to the metastable nature of some phases and the remarkable high power performance of this material. In order to get a better insight into the electrochemical process, a local and fast method of analysis is thus necessary. In positive lithium battery materials, electron diffraction or high resolution transmission electron microscopy (HRTEM) have been used to determine local structures.[4, 6-10] Although these techniques can meet the spatial resolution requirement, they both have very time consuming limitations: electron diffraction necessitates the orientation of each crystal and HRTEM further demands supporting image simulation. Local phases can also be obtained from core losses in electron energy-loss spectroscopy (EELS).[11, 12] The Fe $L_{2,3}$ edge has been used to characterize intercalated/deintercalated phases in $Li_xFePO_4$ grains,[2, 13] and analyze the phase front.[2] Nevertheless, using a Hartree-Slater model, the cross section for the Fe $L_{2,3}$ edge (situated around 708 eV) can be shown to be 30 times smaller than that of the Fe $M_{2,3}$ edge (around 54 eV).[14] Miao *et al.* published a study on that precise edge but the energy resolution was too poor to give any insight on the chemical process.[13] Valence EELS (VEELS), hence losses below 40 eV, is even more interesting since its intensity is approximately 10 times more intense than that of the $M_{2,3}$ edge. Such intensity could help to meet the time scale criteria in order to both perform kinetic studies and limit electron beam damage.

In this letter, after comparing FePO$_4$ and LiFePO$_4$ VEELS spectra with calculations, we show that a particular peak, not sensitive to the precise orientation of the sample, allows the determination of local compositions in Li$_x$FePO$_4$ samples. This same peak is demonstrated to be useful in the fast recording of phase sensitive energy filtered images, a prerequisite to gain insight into the dynamics of the intercalation process.

In order to identify the most relevant features in FePO$_4$ (FP) and LiFePO$_4$ (LFP) VEELS spectra, experiments are compared to first principle calculations in Fig. 1. Calculations were performed with WIEN2k,[15] a full potential code, based on the density functional theory (DFT). In order to take into account local magnetic moments on Fe$^{II}$ (LFP) and Fe$^{III}$ (FP) sites, spin polarized calculations were necessary. An antiferromagnetic (AF) order (AF along the channel **b** direction but ferromagnetic in the two others) was introduced together with the use of GGA+U approximation, considering a mean U$_{eff}$ value (4.3 eV) for both LFP and FP, as previously done by Zhou *et al*.[16] Dielectric functions were obtained in the optic approximation (q = 0).[17] Atomic coordinates were taken from Delacourt *et al*.[18] and details of the calculations are given as supplementary materials.[19] Analyzed samples were synthesized by the nitrate method,[20] so that no carbon coating, which could modify intensities in the VEELS region, was present. Chemically delithiated compounds were obtained by using proper amounts of NO$_2$BF$_4$.[2] VEELS spectra were obtained on a HF2000 Hitachi cold field emission gun TEM operated at 100 kV and at liquid nitrogen temperature to avoid beam damages. The probe size was around 60 nm in diameter, in order to avoid once again beam damages, which are especially large in the case of FP. A modified Gatan 666 spectrometer was used giving a 0.65 eV energy resolution at a 0.05eV/pixel dispersion.[21] VEELS were acquired in less than 0.2 s total time. Experimental spectra were gain, dark count, zero loss (ZL) and plural scattering corrected using the PEELS program.[22] By applying the same method as previously published, a common intensity scale (eV$^{-1}$) is obtained for both experimental and theoretical spectra.[23, 24] Scissors operators corresponding to upwards energy shifts of 1 eV and 2 eV were used for FP and LFP, respectively.[25] For both FP and LFP, experimental and calculated spectra are in very good agreement. Most peaks occur at the same energy with a comparable relative intensity. This proves both the validity of discussions based on theoretical considerations and the actual acquisition of spectra free of beam damages. The main difference between FP and LFP spectra is the 3-eV wide peak situated around 6 eV. The study of total, partial and joint densities of states for FP and LFP are particularly useful to explain the origin of this "6 eV peak".[26] In fact, this peak is also present in LFP but is hidden in the main intensity rise above 7.5 eV. Such a large shift is due to the strong Fe 3d destabilization with the increase of occupancy ($3d^6$ in LFP, $3d^5$ in FP), rather than a simple filling of empty levels via Li intercalation. Further calculations of FP energy-loss functions (ELF) in the x, y and z directions also prove that this peak is always present,[26] and that the study of $Li_xFePO_4$ samples using this peak is essentially independent of the precise crystal orientation.

The first application of EELS measurements of this 6 eV peak is the determination of lithium compositions in single phase $Li_xFePO_4$ samples. As an example, we present experimental [Fig. 2(a)] and theoretical [Fig. 2(b)] spectra for a single phase sample and compare them to the respective FP and LFP spectra. A $Li_{0.6}FePO_4$ powder was obtained by chemical delithiation using a proper amount of $NO_2BF_4$. The single phase sample was obtained by rapid quenching in water of this same $Li_{0.6}FePO_4$ powder from 300°C.[13] In order to simulate an intermediate composition, the geometry of a supercell corresponding to a $Li_{0.5}FePO_4$ composition was optimized using the VASP program.[27] Experimental cell parameters were taken from Ref. 18. Optimized atomic parameters as well as magnetic ordering details for this structure are given as supplementary materials.[19] The same $U_{eff}$ as for LFP was used. The optimized structure was then included in WIEN2k to obtain the dielectric functions for this compound. Since 1 and 2 eV scissors operators were included to fit experiments in Fig. 1, an arbitrary 1.5 eV upward shift for $Li_{0.5}FePO_4$ was used. These results mean that the 6 eV peak is roughly proportional to the lithium content in $Li_xFePO_4$ and can be used for a local quantification of lithium. Considering the latest development of single phase samples for nanograins (d < 40 nm),[28] such a fast and local determination is particularly helpful. The use of this peak is similar to that of the oxygen K edge prepeak with the advantage of a much lower acquisition time.[2]

A second application of the 6 eV peak is the possibility of distinguishing between LFP and FP phases by acquiring energy filtered images. Energy filtered TEM (EFTEM) has long been demonstrated to allow mapping of local physical properties,[29, 30] but is often based on a small change in VEELS spectra. In our case, VEELS spectra for both compounds are very dissimilar below 8 eV. By selecting electrons in the 5 eV range, energy filtered images with a high contrast between intercalated and non intercalated crystals can thus be obtained.

A $Li_{0.85}FePO_4$ sample, synthesized by chemical oxidation of LFP by $NO_2BF_4$, was analyzed using this technique. Images were acquired on a JEOL 2200FS at 200kV equipped with an in-column omega filter. A bright field image [Fig. 3(a)] is compared to an energy filtered image acquired with the energy slit centered at 5 eV [Fig. 3(b)]. The width of the slit was 3 eV with a 10s acquisition time. In Fig. 3(b), some crystals emerge much brighter than others. This contrast is not due to thickness effects. For instance, crystals labeled 1 and 2 present both t/λ close to 0.3. Crystal 2 is dark whereas crystal 1 is lit, illustrating the existence of the 6 eV peak (FP) in its EELS spectrum. VEELS spectra recorded on these two areas indeed confirm that area 2 corresponds to an LFP crystal and area 1 to a FP crystal. In most cases, FP crystals were found on edges of big aggregates, in accordance with the chemical delithiation process used for our sample synthesis.

It is worth noting that EFTEM images were acquired in a few seconds and without post-treatment. A fast and simultaneous determination of phases in multiple crystals was thus possible. The electron beam was spread out over a very large area, therefore limiting beam damages while keeping a spatial resolution close to 10 nm in our case. The EFTEM experiments could thus be performed at 200 kV and at room temperature.

In conclusion, using first principles calculations, the most characteristic features in the low loss spectra of LFP and FP were identified. In particular, the area between 3 and 7 eV was shown to be ideal for a fast, unambiguous and easy determination of the repartition of $Li_xFePO_4$ phases in a sample. This region can be used to quantify the local lithium content as well as produce direct chemical maps of intercalated or non intercalated crystals, independently of their orientation. The quickness of the method opens the way to kinetic studies of the lithium intercalation process in $Li_xFePO_4$ crystals.

ACKNOWLEDGMENTS

We thank J. Gaubicher (IMN) for supplying the starting $LiFePO_4$ material. Computations presented in this work were performed at the "Centre Régional de Calcul Intensif des Pays de la Loire" financed by the French Research Ministry, the "Région Pays de la Loire", and the University of Nantes.

Figure captions:

FIG. 1. (a) VEELS spectra for $FePO_4$. Thin dashed line: experiment; thick line: calculation. (b) VEELS spectra for $LiFePO_4$. Thin dashed line: experiment; thick line: calculation.

FIG. 2. (a) Experimental VEELS spectra. Normalized at 50 eV. (b) calculated VEELS spectra, shifted according to Fig. 1 shifts. Thick line: $LiFePO_4$; thin line with dots: $Li_{0.5}FePO_4$ single phase; thin line: $FePO_4$.

FIG. 3. (a) Bright field image of a region with multiple nanocrystals (b) EFTEM image of the same area for a slit centered at 5 eV.

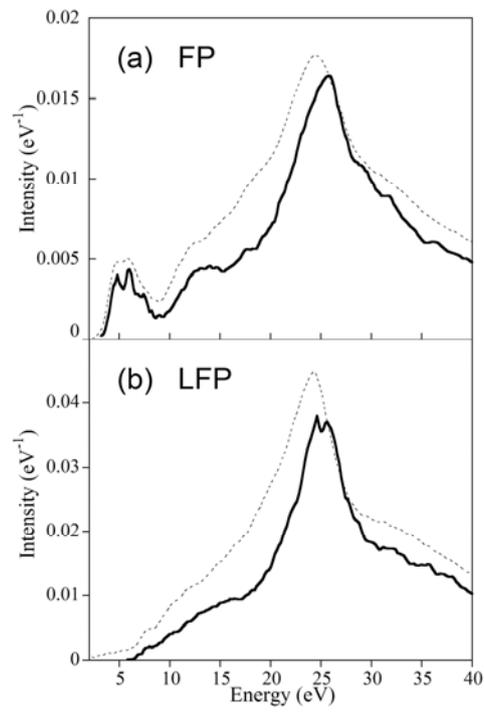

Figure 1

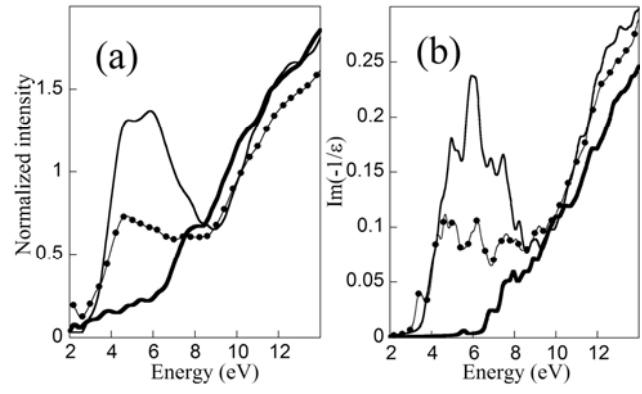

Figure 2

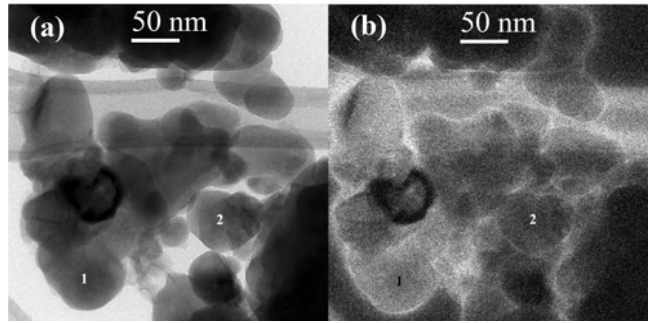

Figure 3